\documentclass{article} % For LaTeX2e
\usepackage{iclr2024_conference,times}

\usepackage{amsmath,amsfonts,bm}

\def\Figref#1{Figure~\ref{#1}}

\def\Secref#1{Section~\ref{#1}}
\def\eqref#1{equation~\ref{#1}}
\def\Eqref#1{Equation~\ref{#1}}
\def\1{\bm{1}}

\DeclareMathAlphabet{\mathsfit}{\encodingdefault}{\sfdefault}{m}{sl}
\SetMathAlphabet{\mathsfit}{bold}{\encodingdefault}{\sfdefault}{bx}{n}

\usepackage{hyperref}
\usepackage{url}

\usepackage{graphicx}
\usepackage{bm}
\usepackage{bbm}
\usepackage{amsmath}
\usepackage{color}
\usepackage{xspace}
\usepackage{amsfonts}
\usepackage{booktabs}
\usepackage{multirow}
\usepackage[ruled]{algorithm2e}
\usepackage{lineno}
\usepackage{wrapfig}
\graphicspath{{./figures/}}

\usepackage{pifont}% http://ctan.org/pkg/pifont
\newcommand{\cmark}{\ding{51}}%
\newcommand{\xmark}{\ding{55}}%

\def\onedot{.\xspace}
\def\eg{\emph{e.g}\onedot} 
\def\ie{\emph{i.e}\onedot} 
 
\def\etc{\emph{etc}\onedot}

\def\fullname{FuzzPretrain\xspace}
\def\abbrname{FuzzPretrain\xspace}
\title{Code Representation Pre-training \\ 
with Complements from Program Executions}

\author{
Jiabo Huang$^{1}$, Jianyu Zhao$^{1}$, Yuyang Rong$^{3}$, Yiwen Guo$^2$\thanks{Corresponding author}\, , Yifeng He$^{3}$,\ Hao Chen$^3$  \\
\small{$^1$Tencent Security Big Data Lab,\, $^2$Independent Researcher,\, $^3$UC Davis}\\
\small{\texttt{\{jiabohuang,yjjyzhao\}@tencent.com\quad \{PeterRong96,guoyiwen89\}@gmail.com}
} \\
\small{\texttt{\{yfhe,chen\}@ucdavis.edu}
}}

\iclrfinalcopy % Uncomment for camera-ready version, but NOT for submission.
\begin{document}

\maketitle

\begin{abstract}
% The successes attained by 
Large language models 
(LLMs) 
for natural language processing % in recent years
have been grafted onto %constantly being grafted onto 
programming language modeling
for advancing code intelligence.
Although it can be represented in the text format, 
code is syntactically more rigorous
in order to be properly compiled or interpreted
to perform a desired set of behaviors
given any inputs.
% % challenges
In this case, existing works
benefit from syntactic representations
% to learn from the structure of code
to learn from code
less ambiguously
% by presenting its structure
% by learning from its 
in the forms of
% as
abstract syntax tree, control-flow graph, \etc
% limitations
However,
programs with the same purpose can be
implemented in various ways 
showing different syntactic representations
% while code with similar structures
% while code looks similar
while the ones with similar implementations
can have distinct behaviors.
Though trivially demonstrated during executions, such 
% implementation variations
% functionality-agnostic variations
semantics about functionality
% functional semantics
are challenging to be learned directly from code, %structures, 
% and other syntactic representations,
especially in an unsupervised manner.
Hence, in this paper,
we propose \fullname
to explore the dynamic information of programs
revealed by their test cases
% as functional representations of code
and embed it into the feature representations of code
as complements.
% to complement the static information derived from code or its syntactic representations.
%from the structures.
% and syntactic representations.
% to complement previous code representations by
% exploring
% the dynamic information of programs
% revealed by their
% input-output representations (\ie, test cases).
% The inputs are synthesised efficiently by fuzzing
% while the outputs are yielded by executions.
% Both of which are only required for pre-training.
The test cases are obtained 
with the assistance of 
a customized fuzzer
and are only required during pre-training.
%
% The FuzzPretrain models yielded
FuzzPretrain
yielded more than $6\%/19\%$ 
mAP improvements
on code search
over its counterparts trained with
only source code or AST, respectively.
Our extensive experimental results
show the benefits of 
learning discriminative code representations with % complements from 
program executions.
% Through extensive experiments on 
% various code understanding benchmarks,
% % code understanding and generation benchmarks,
% we demonstrate, for the first time, that the advocated
% dynamic program information
% is indeed beneficial to learning 
% discriminative code representations
% % when only source code
% % is usually available 
% in practice.
% and help with various downstream tasks
% when only the code of programs is available.
\end{abstract}

\section{Introduction}

\begin{figure*}[t]
\begin{center}
\includegraphics[width=1.0\linewidth]{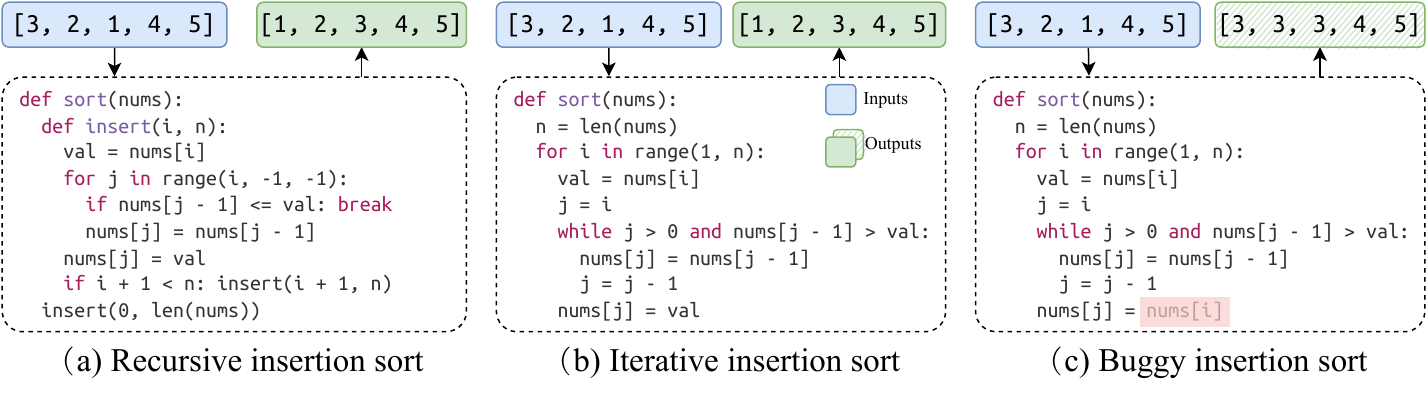}
\end{center} \vskip -0.2in
\caption{
An illustration of 
implementation variations of the same functional purposes.
The source code of
\textbf{(a)} recursive insertion sort
is dramatically different from its
\textbf{(b)} iterative counterpart regarding their structures,
even though their functional equivalence
is explicitly demonstrated
by the consistent behaviors.
On the other hand,
the subtle change in
\textbf{(c)} %the buggy implementation
% is unobservable 
is buggy and barely observable
in comparisons to (b), 
but can lead to distinct execution results.
% or even software vulnerabilities
% that have catastrophic results.
} \vskip -0.2in
\label{fig:teaser}
\end{figure*}

Code representation learning %~\citep{feng-etal-2020-codebert,guo2020graphcodebert}
is drawing growing attention
across the community of
artificial intelligence (AI) and 
software engineering (SE)~\citep{husain2019codesearchnet,deng2023large,liu2023code}.
It aims to 
% The objective is to
abstract 
the structure and the underlying functionality of the source code
% according to their code
and embed such semantics into a latent representational space via unsupervised pre-training.
% It plays a significant role 
By providing general understanding of programs, 
it is a fundamental task %yet challenging task
to achieve code intelligence, 
which enables automated code analysis and processing %tasks
% to facilitate various code-related downstream tasks
by fine-tuning with budgeted computation resources or
limited human annotations~\citep{rattan2013clone_detection,husain2019codesearchnet,lu2021codexglue}.
The pre-training recipes~\citep{devlin-etal-2019-bert,liu2019roberta} for natural languages
have been shown effective in code representation learning~\citep{feng-etal-2020-codebert,radford2019gpt2}.
However,
they neglect that
% the compositions of code 
the structure of the code 
(exhibited by how its elements, \eg, variables and statements, are organized)
must comply with precise and rigorous rules.
These rules, often referred to as code syntax, are
% The syntax is
defined by language specification
to ensure successful compilation and execution.
This inspires
% is highly structured and contains much richer information than natural language. Therefore, several 
recent works to %on code representation learning
leverage static analysis~\citep{wichmann1995industrial}
% to present its structure less ambiguously by
to parse and 
present the structure of code less ambiguously by
% more definite 
its syntactic representations,
% parsed from source code,
\eg abstract syntax tree (AST)~\citep{guo2022unixcoder,tipirneni2022structcoder} 
and control-flow graph (CFG)~\citep{allamanis2017learning}.
However,
since learning from code as it is a type of static data,
existing methods overlook the underlying executable programs.
Whilst
programs are built with specific purposes
of performing a set of behaviors
% which is indicated by their functionality,
indicated by their functionality,
it is challenging
to derive such semantics
% from code and its syntactic representations~\citep{liu2023code}.
from code structure~\citep{liu2023code}.
This is because the same purposes
can be implemented by different algorithms in different ways,
while the behaviors of programs are susceptible to subtle discrepancies in code.
% This is because the algorithms expressed in programming languages are rigorous while the same problems can be solved by different algorithms and different implementations.
% the variability of programming languages allows
% large ``intra-behavior'' implementation discrepancies
% and ``inter-behavior'' similarities.
% However,
% it is inherently challenging to derive the functionality
% of programs according to only their static descriptions
% without explicit indications of
% their logic paths.
% % which maps the inputs into outputs.
% Meanwhile,
% this is critical for program comprehensions
% in lots of code intelligence applications
% like identifying functionally equivalent code snippets
% in a large-scale software project
% to refrain from the costs of maintaining their consistent behavior.
% Concretely,
As depicted in \Figref{fig:teaser} (a) and (b),
the recursive and iterative %implementations of 
insertion sort
are notably different regarding their structures, 
even though they share the same functionality.
% and always behave the same.
% (intra-behavior discrepancy).
Meanwhile,
the subtle change in \Figref{fig:teaser} (c)
that is hard to be noticed can lead to distinct execution results.
% (inter-behavior similarity).
% Therefore,
How to learn discriminative feature representations of code
that embed not only
the static information from its structure
but also the dynamic information about its functionality
% \textit{structural semantics}
% about its compositions and organisations
% but also the \textit{functional semantics} %of the underlying programs
% about its purposes and behavior
% remains an open question.
remains an unsolved problem.

In this work, 
we aim to embed the functional purposes of code into its feature representations,
% to address the limitations of previous works
% when learning from code or its syntactic representations solely 
% and overlooking their potential variations
% that are insensitive to functionality.
to address the aforementioned limitations of existing works.
The key idea is to 
abstract the behaviors %and purposes 
of programs
% abstract program's behaviors and purposes %of programs
from their input-output relationships 
(represented by test cases) 
and enforce the model to infer 
such information from source code %or its syntactic representations 
that are readily available in downstream tasks.
% 
% To obtain sufficient test cases, 
To achieve this, we take advantage of fuzzing~\citep{zeller2019fuzzing}
% which is capable of generating 
to produce
test cases that cover the logic paths of code as comprehensive as possible.
Moreover,
we propose a novel method called \textbf{\fullname}
for joint static and dynamic information modeling.
% we adopt an off-the-shelf fuzzing approach,
% which is well-known to be a primary dynamic analysis technique,
% to generate diverse input-output pairs (test cases) for programs
%
% Specifically,
% beyond learning the static program information
% by the conventional masked language modeling objective,
% the proposed \abbrname model is optimised to accomplish
% In particular,
Particularly,
in addition to exploring code structure by masked language modeling~\citep{devlin2018bert},
% beyond adopting the conventional masked language modeling~\citep{devlin2018bert} for structure exploration,
it formulates a dynamic information matching (DIM) pretext task
to tell test cases of different programs apart
according to their correspondence to code.
By doing so,
the model learns holistic feature representations of code
% as well as its test cases which embed the program-specific functionality.
and its test cases, encoding 
% semantics about 
both the structure and functionality.
% the model will learn to derive the input-output mappings 
% embedding the program-specific functionality.
% so as to derive the program-specific runtime behavior
% from test cases and help with understanding the code.
% The proposed novel method, called \textbf{\fullname}, is also formulated
\abbrname also involves a self-distillation objective
to accomplish a dynamic information distillation (DID) objective.
% This is for learning the feature representations of code by 
% distilling
% from the holistic representations.
% encoding both the static and dynamic information
% when they are both available in pre-training.
Thereby,
the dynamic information is not only properly modelled
but distilled from the holistic representations
to code features,
so to benefit in practice
where the test cases are 
% inaccessible and 
not required.

We make three contributions in this paper:
\textbf{(1)}
We propose to leverage the test cases of programs
obtained with the help of fuzzing
as explicit indications of functionality
to complement their code and syntactic representations.
To the best of our knowledge,
% To our best knowledge,
this is the first attempt
to unveil the benefits of dynamic program information
on code representation pre-training.
\textbf{(2)}
We introduce a novel code pre-training method named \abbrname.
It simultaneously 
models the structure and functionality of code
while distilling from such holistic information 
to represent code in its feature space.
% to code feature representations.
% models the static and dynamic information of programs
% while distilling from their holistic representations
% to be encoded into code feature representations.
% when only the source code is available
It is ready to benefit %code-related
downstream tasks 
without extra cost on test cases generations.
\textbf{(3)}
% We construct
% a large-scale, diverse, and comprehensive code corpus
% with each program is described in both the 
% static (code) and the dynamic (test cases) forms
% to inspire future studies of the field.
Extensive experiments on four code understanding downstream tasks
% and five datasets
demonstrate the effectiveness of \abbrname
on complementing both source code and its syntactic representations, \eg AST,
by dynamic program information 
for learning discriminative feature representations.

\section{Related Work}
% In this section,
% we summarise how code intelligence~\citep{feng-etal-2020-codebert}
% is promoted by the developments of LLMs~\citep{devlin2018bert}
% in recent years.
% Then,
% we briefly introduce
% the code analysis techniques~\citep{fioraldi2020afl_pp} widely adopted
% in software testing as well as
% their intersection with AI~\citep{deng2023large,liu2023code,zhao2023understanding}.

\paragraph{Code representation learning.} %by large language models.}
Large language models~\citep{raffel2020exploring,liu2019roberta,devlin2018bert}
have achieved unprecedented breakthrough in natural language processing in recent years~\citep{vaswani2017attention,devlin2018bert,radford2019gpt2}.
% thanks to the superiority of Transformer~\citep{vaswani2017attention} 
% on sequence analysis
% and 
% effective self-supervised pretext tasks~\citep{devlin2018bert,radford2019gpt2}
% % for unsupervised learning.
% for unsupervised learning.
Such successes of LLMs have been consistently transferred
to code representation learning and advance code intelligence.
Early works in the field 
devote to building large-scale %and diverse 
code corpus~\citep{puri2021codenet,husain2019codesearchnet}
to be trained simply as plain text
at the natural language conventions~\citep{feng-etal-2020-codebert,mark2017codex,lu2021codexglue}.
% \jz{However, the structures of code provides crucial code semantics and would enhance the code understanding process.}
However,
the rigorous syntax of programming languages
exhibit additional information
about code semantics~\citep{hellendoorn2019global,guo2020graphcodebert}.
In this case,
recent efforts %in this field 
turn to code-specific designs,
\eg, pre-training by
identifier detection and infilling~\citep{wang2021codet5}
% or learning from syntactic representations~\citep{guo2022unixcoder,allamanis2017learning}.
or parsing the structure of code by its syntax~\citep{guo2022unixcoder,allamanis2017learning}.
Despite being effective,
existing approaches %for code representation learning
consider code as a type of static data and ignore the fact that
it is corresponding to an executable program with unique functionality
exhibited by its runtime behaviors.
% that may vary in different runtimes.
Such dynamic information of programs
is a critical indicator to tell them apart from others with different purposes
but challenging to be inferred 
% implicitly 
from code structures.

\paragraph{Static and dynamic code analysis.}
Static and dynamic analysis are 
two common strategies in SE for
ensuring the security of software products, 
and the former has been widely adopted in 
code representation learning for 
constructing syntactic representations~\citep{guo2022unixcoder,guo2020graphcodebert}.
% Whilst static analysis has been widely adopted
% in exiting code representation learning methods
% for constructing syntactic representations,
% it and dynamic analysis are 
% two commonly used strategies in SE
% to ensure the security of software products. 
For early detection of potential vulnerabilities, %and bugs,
static analysis~\citep{wichmann1995industrial} 
is usually carried out
by parsing the structure information of the code for inspections,
while dynamic analysis~\citep{khatiwada2018just}
detects runtime errors during program executions.
Fuzz testing~\citep{zeller2019fuzzing}, or fuzzing,
is a popular dynamic analysis technique, %for dynamic analysis,
which aims to generate a set of inputs that achieve high code coverage
then feed those inputs to programs for execution to observe any unexpected outcome.
Overall,
the two types of code analysis strategies 
are usually adopted together as mutual complements
to ensure the soundness and completeness of testing.

\paragraph{Large language models meet software testing.}
% As LLMs exhibited impressive capabilities of code generation,
There are 
recent efforts on
automated bugs mining by LLMs~\citep{schafer2023adaptive,kang2022large},
which hold an opposite objective to ours
on benefiting software testing by code generation.
% on benefiting code comprehensions by software testing techniques.
% Their primary objective is on vulnerabilities detection 
% while ours is on comprehensive programs profiling.
On the other hand,
\citet{guo2020graphcodebert} demonstrate
the effectiveness of code understanding
from data-flow graph (DFG)
extracted via static analysis.
Although the intricate dependencies among variables somehow imply
the functionality,
it suffers from the same problems
as the methods introduced by~\citet{liu2023code,zhao2023understanding},
\ie collecting additional information
about structure or functionality of code
requires sufficient expertise in SE
and this undoubtedly hampers model's applicability
when it is required on every downstream task.
% it suffers from the same problems
% as the studies performed by
% \citet{liu2023code} and \citet{zhao2023understanding}, which
% show that execution traces and test cases 
% are helpful %to code understanding
% when they are available as additional information.
% Collecting such information about 
% the structure or functionality of code
% is non-trivial and requires sufficient expertise in SE.
% This undoubtedly restricts the deployability of their learned models in practice. %of those methods in practice 
By contrast,
we aim to explore program executions only in pre-training
to preserve the benefits of dynamic information to code understanding
in practice where test cases are not mandatory.
% on downstream tasks where test cases are not required.
% without additional costs on fuzzing.
% without additional costs and efforts on 
% test cases compositions, 
% programs compilations or executions.

\section{Code Representation Pre-training}

\begin{figure*}[t]
\begin{center}
\includegraphics[width=1.0\linewidth]{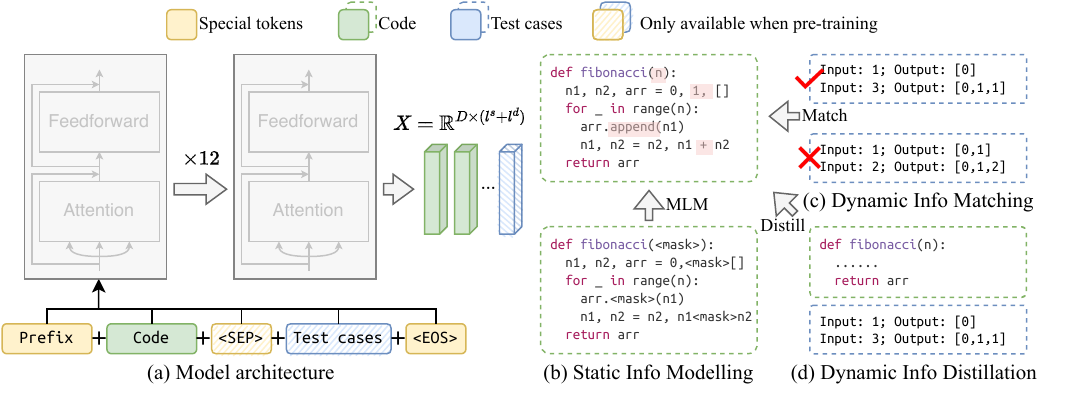}
\end{center} \vskip -0.2in
\caption{An overview of \fullname.
\textbf{(a)} The input sequences %to \abbrname
are composed of both the code 
and test cases
which are concatenated then encoded by a transformer.
% \abbrname learns code representations
% by accomplishing three pretext tasks including,
% \textbf{(b)} static information modeling (SIM)
% through masked tokens predictions
% to explore the structure of code,
% \textbf{(c)} dynamic information matching (DIM)
% to summarize the execution behaviors of programs
% from their test cases
% and 
% \textbf{(d)} dynamic information distillation (DID)
% so that the code representations are enforced to be
% aware of both the static and dynamic information
% even when test cases are not available on downstream tasks.
\abbrname learns code feature representations
by
\textbf{(b)} static information modeling (SIM)
through masked tokens predictions,
\textbf{(c)} dynamic information matching (DIM)
to match test cases to code, and
\textbf{(d)} dynamic information distillation (DID)
to summarize the holistic information about code structure and functionality.
} \vskip -0.1in
\label{fig:pipeline}
\end{figure*}

Given a piece of code $T^s=\{\bm{t}^s_i\}_{i=1}^{l^s}$
composed of $l^s$ tokens % $\bm{t}^s_i$
and a sequence encoder $f_\theta$ 
% (\eg a transformer~\citep{vaswani2017attention})
parameterized by $\theta$,
our objective
% the objective of code representation learning
is to explore the underlying semantics of the code $T^s$
% \eg its structure and functionality,
% and encode such semantics 
and encode them in a latent representational space
$X^s=\{\bm{x}_i^s\}_{i=1}^{l^s}=f_\theta(T^s)\in \mathbb{R}^{D\times l^s}$
in $D$-dimensions.
This is to provide general understanding of code,
which enables efficient fine-tuning on downstream tasks.
% The learned feature representations 
% embeds general understanding of code, which 
% can then be used to be fine-tuned on various
% code-related downstream tasks efficiently. %regarding both time and data.
% Although programming languages are rigorous in syntax,
% the functional purposes of programs are 
% hidden in the intricate logic paths of code
% and hard to be learned accurately from its structures.
% Therefore,
% comprehensive program understanding is fundamentally challenging
% to facilitate effective code representation learning.

In this paper,
we propose to explore the dynamic information obtained during program executions
to complement the static information learned from code structure,
such that we can embed both in feature representations of code.
To that end,
we formulate \textbf{\fullname} %for code representation learning,
whose overview is depicted in \Figref{fig:pipeline}.
We first collect a large-scale code corpus based on CodeNet~\citep{puri2021codenet}
and pair each code snippet with multiple test cases
% $T^d_i=\{\bm{t}^d_{i,j}\}_{j=1}^{l^d_i}$
synthesized with the assistance of a customized fuzzer.
% For clarity,
We denote 
% the whole list of 
the %concatenated
test cases corresponding to $T^s$ 
as $T^d=\{\bm{t}^d_i\}_{i=1}^{l^d}$, which is
composed of $l^d$ tokens.
The \fullname
then concatenates the test cases
with the code as $T=\{T^s \oplus T^d\} = \{\bm{t}_i\}_{i=1}^{l}$
where $l=l^s+l^d$.
% holding the assumption that
% the input-output relationships are
% embedding the functionality of programs
% and likely beneficial to the comprehension of code.
% For clarity,
% We denote the length of the concatenated sequence $T$ to be $l=l^s+l^d$.
By feeding $T^s$ (or $T$) into $f_\theta$,
the features $X^s$ (or $X$) are trained by 
%the following pretext tasks.
% The static
% and dynamic information modeling
% are carried out by 
masked tokens predictions (\Figref{fig:pipeline} (b))
and test cases to code matching (\Figref{fig:pipeline} (c)).
% and matching test cases to code (\Figref{fig:pipeline} (c)),
% respectively.
% static information modeling (SIM) (\Figref{fig:pipeline} (b))
% for masked tokens predictions and
% dynamic information matching (DIM) (\Figref{fig:pipeline} (c))
% to derive the behavior from input-output relationships embedded in test cases.
% Besides,
% \abbrname formulates a
% dynamic information distillation (DID) (\Figref{fig:pipeline} (d)) objective
% % to distil the rich knowledge about programs from $T$
% to distill %the dynamic information 
% from the holistic features $X$
% % into $X^s$ so to benefit downstream tasks when 
% % only the code $T^s$ is usually available.
% into $X^s$ in order to adapt to downstream tasks
% where test cases $T^d$ are not available.
% % the functional representation $T^d$ is not available. 
Besides,
\abbrname 
distills from the holistic features $X$ of code and test cases
and embed it into $X^s$,
in order to adapt to downstream tasks
where test cases $T^d$ are not available.

\subsection{Fuzzing Code Corpus}

% \paragraph{Fuzzing for test cases}
% Fuzzing, a widely embraced input-centric technique for dynamic program analysis, plays an important role in identifying vulnerabilities and enhancing software reliability.
Fuzzing is a software verification technique that plays an important role in identifying vulnerabilities and enhancing software reliability.
% The fuzzing procedure can be described as a favored input modification process, where the fuzzer initially executes the program with a given set of inputs. Subsequently, the behavior monitor prioritizes inputs that traverse uncharted paths within the program's execution tree. 
% A fuzzer verifies the software by recently generating inputs for the software to execute.
A fuzzer verifies the software by repeatedly generating inputs for the software to execute.
For each execution, the fuzzer monitors the internal state of the software to determine if the input triggers new behavior. These inputs will be stored for future input generation.
% This step enables the fuzzer to effectively focus on exploring unexplored program behaviors. 
Input generation and behavior monitor together allow the fuzzer to effectively focus on exploring new program behaviors.
% Finally, the input mutator employs diverse strategies to modify the inputs according to their priority, facilitating the initiation of another iteration of the fuzzing loop. 
We believe these inputs contain runtime information that cannot be easily found using static analysis.
Therefore, using those test cases, \ie, program inputs and corresponding outputs, should supply extra dynamic information to the language model.
%In our endeavor to obtain dynamic program information, we employ the methodology outlined in FuzzTuning \citep{zhao2023understanding} to carry out preprocessing, compilation, and fuzzing of the code corpus.
We employ methods outlined in FuzzTuning \citep{zhao2023understanding} to carry out preprocessing, compilation, and fuzzing of the code corpus.
In brief,
given a C++ or Java code snippet,
% we first [preprocess (compliation?)]
we use a customized compiler to instrument it before compiling it to an executable.
For Python, we modified the interpreter to report the program behaviors during execution.
% then [fuzz]
We fuzz the program using AFL++ \citep{fioraldi2020afl_pp} and extract the inputs.
% and [execute].
Finally, we re-run the program and record the output of the execution.
More details about fuzzing existing code corpus can be found in \citet{zhao2023understanding}.

\subsection{Static and Dynamic Information modeling}
% The dynamic information obtained from program executions
% is potentially complementary to 
% not only the source code but also its syntactic representations.
To investigate the versatility
of learning with 
dynamic program information,
% obtained from program executions, 
% to different 
% static-based methods,
we build the \abbrname
upon two representative models
trained on either code or AST,
namely CodeBERT~\citep{feng-etal-2020-codebert}
and UniXcoder~\citep{guo2022unixcoder}, respectively.
% Both the two base models 
% follow \citet{liu2019roberta}
% to encode input sequences by a transformer encoder~\citep{vaswani2017attention}.
% but the prefix-adaptive design of UniXcoder
% enables it to be deployed 
% on both discrimination and generation tasks.
% Note that,
We want to emphasize that
\abbrname is a generic method
that can be integrated into arbitrary static-based models
more than the two studied here.
\abbrname is ready to benefit different presentations of code 
in a plug-in manner
once they are serialized as a sequence of tokens.
For clarity,
we introduce our designs %of \abbrname
in terms of source code inputs in this section.
% They are directly applicable to AST used by UniXcoder
The designs for other models like UniXcoder are similar.
% which is directly transferable to 
% AST in UniXcoder~\citep{guo2022unixcoder}
% as being flattened.
% for simplicity,
% we introduce \abbrname in terms of code inputs
% in the following sections,
% which is directly transferable to AST
% % or any other syntactic representation
% when being flattened and learned
% as a sequence of tokens~\citep{guo2022unixcoder}.

\paragraph{Input/Output representations.}
As illustrated in \Figref{fig:pipeline} (a),
we follow \cite{feng-etal-2020-codebert}
to concatenate different parts of inputs together with
an \texttt{<SEP>} token
% then prepend model-specific prefix to the inputs with 
and put an end-of-sentence \texttt{<EOS>} token to the end of the concatenation.
% For the model-specific prefix,
CodeBERT adopts a begin-of-sentence token \texttt{<BOS>}
as the prefix for the input sequences
while UniXcoder takes \texttt{<BOS><ENCODER-ONLY><SEP>}~\citep{guo2022unixcoder}.
% We denote the \texttt{[Prefix]} 
% in \Figref{fig:pipeline} (a)
% using square brackets
% to indicate it is not a specific token in our vocabulary.
For the test cases,
we follow \citet{zhao2023understanding}
to decode them from a series of bytes to Unicode strings
and then prompt them in a form of natural language to be
``Input is: \texttt{INPUT}; Output is: \texttt{OUTPUT}''.
% To be specific,
% for a test case
% composed of \texttt{INPUT} and \texttt{OUTPUT},
% we prompt them as
% ``Input is: \texttt{INPUT}; Output is: \texttt{OUTPUT}''.
This is inspired by
how programming online judgement tools 
(\eg Leetcode\footnote{An example of programming online judgement tools: https://leetcode.com})
present problem descriptions
along with its example test cases to human.
We concatenate multiple test cases of a program
with the \texttt{<SEP>} token to form $T^d$.
We learn from multiple test cases at a time
because
a single test case is likely to invoke only a part of a program, and only with sufficient number of test cases 
can we profile the behaviors of the program comprehensively.

We follow the common recipe of natural language processing to
split the concatenation of prefix, code, test cases, and suffix as WordPiece~\citep{wu2016google}.
% to obtain the input sequence
% $T=\{\bm{t}_i\}_{i=1}^{l}$.
We omit the tokens of prefix and suffix in notations
to be unified for both the base models.
By feeding $T^s$ into the sequence encoder $f_\theta$,
we can compute the feature of each token $\bm{x}^s_i$, \ie, we have 
$X^s=\{\bm{x}^s_i\}_{i=1}^l\in\mathbb{R}^{D\times l^s}$.
Given the token-wise representations,
CodeBERT adopts the feature of the \texttt{<BOS>} token
as the code-level representation $\bm{x}^s$
while UniXcoder applies average pooling on all the tokens in a sequence
to obtain $\bm{x}^s$.

\paragraph{Static Information modeling.}
To learn from the structure of code $T^s$
according to the dependencies among tokens,
% To model the structure of the static manifestation $T^s$ of a program,
we adopt the conventional masked language modeling (MLM)
which has been shown simple yet effective
on context understanding~\citep{devlin2018bert}.
We follow the common practices to
randomly choose $15\%$ of the tokens in $T^s$
and replace $80\%$ of the selections with
a special \texttt{<MASK>} token,
$10\%$ with random tokens %from the vocabulary
and the remaining %$10\%$ 
are left unchanged.
% \jz{replace 10\% of them  with random word, and the rest 10\% unchanged}
Formally,
given the code $T^s$,
a subset $T^m \subset T^s$ of it is masked out
and leaving a sequence $\tilde{T}^s$ with replaced tokens.
Then,
the learning objective is to predict $T^m$ given the context in $\tilde{T}^s$:
\begin{equation}
\mathcal{L}_\text{SIM}(T^s) = -\sum_{\bm{x}_i\in X^m} \log p(\bm{x}_i\vert \tilde{X}^s),
\label{eq:sim}
\end{equation}
where $X^m$ and $\tilde{X}^s$ are the features 
obtained from
the outputs of $f_\theta$,
given the tokens being masked $T^m$
and the remaining tokens $\tilde{T}^s$ as inputs,
respectively.
% computed by $f_\theta$ according to 
% the tokens being masked $T^m$ 
% and the input sequence with incomplete context $\tilde{T}^s$,
% respectively.
The term $p(\bm{x}_i\vert \tilde{X}^s)$ denotes the probability that
a replaced token $\bm{x}_i$ from $X^m$ is correctly reconstructed,
given $\tilde{X}^s$.

\paragraph{Dynamic Information modeling.}
To learn from the dynamic program information obtained during executions,
we propose to 
match the input-output mappings derived from test cases
with the functionality inferred from the code.
% Specifically,
Given a code sequence $T^s$,
we randomly sample an unmatched test cases list $T^{d-}$
and decide whether to concatenate $T^s$ with its own test cases $T^d$
or the negative one $T^{d-}$ 
to form the input sequence $T$
at each training step.
We then pair $T$ with a binary label $y\in\{0,1\}$
indicating whether the mapping relationships embedded in the test cases
is consistent with the functionality of $T^s$.
After that,
we feed $T$ to the backbone encoder
followed by an additional linear projection layer and a binary classifier:
% activated by the sigmoid function:
\begin{equation}
\begin{gathered}
\tilde{\bm{x}} = \text{FC}(g(f_\theta(T))),\quad 
p(y\vert\tilde{\bm{x}}) = f_\phi(\tilde{\bm{x}}) \in (0, 1), \\
\mathcal{L}_\text{DIM}(T^s, T^d) = -y\log p(y\vert\tilde{\bm{x}}) - (1-y)\log(1-p(y\vert\tilde{\bm{x}})).
\end{gathered}
\label{eq:dim}
\end{equation}
In \Eqref{eq:dim},
$g(\cdot)$ stands for the operation adopted by the base models 
for aggregating token-level features.
The output of the function $g$
is then linearly transformed to be $\tilde{\bm{x}}$
and classified by $f_\phi$.
The $p(y\vert\tilde{\bm{x}})$ yielded by the classifier $f_\phi$ 
% which is then linearly transformed to be $\tilde{\bm{x}}$.
% The feature $\tilde{\bm{x}}$ is passed to the binary classifier $f_\phi$
% to obtain $p(y\vert\tilde{\bm{x}})$, which 
indicates how likely the code and test cases
in the input sequence $T$ are matched
and is supervised by a binary cross-entropy loss according to $y$.

With the supervision of $\mathcal{L}_\text{DIM}$,
the model is able to derive dynamic information about the execution behaviors of programs and code,
given their test cases.
However,
% such functional representations of code %about the runtime behavior of programs
% are not always available in practice.
% the test cases of programs
test cases
are not available in many downstream tasks %at scale in practice.
in practice.
Therefore,
we further devise a dynamic information distillation (DID)
objective to 
simultaneously 
learn the holistic information from both code and test cases 
$T=\{T^s\oplus T^d\}$
and enforce encoding such information
in the features of code $T^s$.
% enforce the model to 
% preserve the holistic information from both code and test cases
% $T=\{T^s\oplus T^d\}$
% and encode it into the feature representation of code $T^s$.
%
Inspired by \citet{tian2019contrastive},
we formulate DIM in the contrastive learning paradigm
to identify the holistic feature $\bm{x}$
from a list of random samples $X^-$
according to the feature $\bm{x}^s$ of the code $T^s$.
To be concrete,
we follow \citet{he2020momentum}
to maintain a stale copy $f_{\hat{\theta}}$ of the backbone encoder,
which shares the identical architecture with $f_\theta$ and is updated accordingly
% which shares the identical architecture with $f_\theta$
% and is updated at each training step
by exponential moving average (EMA)~\citep{lucas1990exponentially}.
We feed the code $T^s$ into $f_\theta$
to compute its representation, $\bm{x}^s=g(f_\theta(T^s))$
while its holistic counterpart $\hat{\bm{x}}$ is produced by $g(f_{\hat{\theta}}(T))$.
% is encoded by $f_{\hat{\theta}}$ as $\hat{\bm{x}}$.
Given the features of $l^n$ random samples $X^-\in\mathbb{R}^{D\times l^n}$
which are likely with different semantics from $T$,
we train $f_\theta$ to optimize:
\begin{equation}
\mathcal{L}_\text{DID}(T^s, T^s\oplus T^d) = -\log
\frac{\text{exp}(h(\hat{\bm{x}},\bm{x}^s) / t)}
{\text{exp}(h(\hat{\bm{x}},\bm{x}^s) / t) + \sum_{\bm{x}^- \in X^-}\text{exp}(h(\bm{x}^s,\bm{x}^-)/t)}.
% \begin{gathered}
% \mathcal{L}_\text{DID}(T^s, T^d) = -\log p(\hat{\bm{x}}, \bm{x}^s),\ \text{where} \\
% p(\hat{\bm{x}}, \bm{x}^s) = 
% \frac{\text{exp}(\hat{\bm{x}}\cdot\bm{x}^s / t)}
% {\text{exp}(\hat{\bm{x}}\cdot\bm{x}^s / t) + \sum_{\bm{x}^- \in X^-}\text{exp}(\bm{x}^s\cdot\bm{x}^-/t)}.
% \end{gathered}
\label{eq:did}
\end{equation}
The function
$h(\cdot, \cdot)$ in \Eqref{eq:did}
computes the cosine similarity between two vectors, and $t$ is a temperature hyperparameter
controlling the concentration degree of the similarity distribution.
In contrast to $\mathcal{L}_\text{DIM}$,
we always compute the holistic feature $\hat{\bm{x}}$
of code and its own (matching) test cases
to avoid the distractions from inconsistent structure and functionality.
% Such an asymmetric design
% also helps with refraining $f_\theta$ 
% from overfitting to test cases
% which will inevitably harm the generalization ability of the model
% when only code $T^s$ is available in most of the practical scenarios.

\textbf{Remarks.}
Intuitively,
\abbrname can learn dynamic information
by masked token prediction of the concatenation
of code and test cases $T$.
However,
our experiment showed that
the test cases
can be accurately reconstructed in MLM
even without the corresponding code in context.
We believe that the more valuable information
about test cases is their mapping relationships
between inputs and outputs,
instead of the dependencies
between arbitrary tokens.
MLM tends to enforce our model
to memorize every token in test cases,
which is prone to involving the randomness introduced by fuzzing.
This justifies our approach
to derive the potential functionality of programs
% to formulate the dynamic information modeling
according to the correspondence of code and test cases.
More detailed discussions
can be found in \Secref{ssec:analysis}.

\subsection{Model Training and Inference}
% Given the learning objectives of
% static information modeling $\mathcal{L}_\text{SIM}$ (\Eqref{eq:sim}),
% dynamic information matching $\mathcal{L}_\text{DIM}$ (\Eqref{eq:dim})
% and dynamic information distillation $\mathcal{L}_\text{DID}$ (\Eqref{eq:did}),
The proposed \abbrname model is optimized alternatively 
according to
static information modeling $\mathcal{L}_\text{SIM}$ (\Eqref{eq:sim}),
dynamic information matching $\mathcal{L}_\text{DIM}$ (\Eqref{eq:dim}) and
dynamic information distillation $\mathcal{L}_\text{DID}$ (\Eqref{eq:did})
on each mini-batch of data
following~\citep{lample2019xlm,guo2022unixcoder}.
At each training step,
the stale encoder $f_{\hat{\theta}}$ %of the backbone encoder
is updated according to $f_\theta$
by EMA:
$\hat{\theta} = m\hat{\theta}+(1-m)\theta$ with a momentum factor $m$,
and the holistic representations $\hat{\bm{x}}$ %yielded by $f_{\hat{\theta}}$ 
obtained from code and its corresponding test cases
will be fed into the queue $X^-$
with the oldest ones inside being removed in a first-in-first-out manner.
After pre-training, %of code representations,
we keep only the transformer encoder $f_\theta$
which is able to yield discriminative feature representations of code
$X^s=f_\theta(T^s)$
when only it is available but not the test cases $T^d$ at inference or
on downstream tasks.
% Alg.~\ref{alg:fuzz_pretrain} summarises the training process of \abbrname.
% \begin{algorithm}[ht]
% \caption{\abbrname for code representation learning.}\label{alg:fuzz_pretrain}
% \BlankLine
% \KwIn{Fuzzed code corpus, training iterations $N_{it}$.}
% \KwOut{A code-aware language model;}
% \For{$\text{iter}\!=\!1$ \KwTo $N_\text{it}$}{
%     Samples a random mini-batch $B$ of code snippets\;
%     \For{$T^p \in B$}{
%         Samples a random set of test cases $T^f$ for $T^p$\;
%         Concatenates $T^p$ and $T^f$ as $T$\;
%         \tcp{Masked Language modeling}
%         Masks out $15\%$ of the tokens in $T$\;
%         Predicts masked tokens (Eq.~\eqref{eq:loss_mlm})\;
%         \tcp{Contrastive Learning}
%         Updates $f_{\hat{\theta}}$ by EMA (Eq.~\eqref{eq:ema})\;
%         Encodes $T^p/T$ by $f_\theta/f_{\hat{\theta}}$ as $\bm{x}/\hat{\bm{x}}$\;
%         Identifies $\hat{\bm{x}}$ from $X^-$ by $\bm{x}$ (Eq.~\eqref{eq:loss_cl})\;
%         Updates $X^-$ by $\hat{\bm{x}}$\;
%         Updates model weights by SGD\;
%     }
% }
% \end{algorithm}

\section{Experiments}
We collected 1.2M code snippets
% from the submissions
% to $1000/1400/800/250$ coding problems
implemented in C/C++/Python/Java
from CodeNet~\citep{puri2021codenet}
to be fuzzed as the training data of \abbrname.
We followed the base models,
\ie, CodeBERT~\citep{feng-etal-2020-codebert} and UniXcoder~\citep{guo2022unixcoder}
to take a 12-layers transformer
with 125M learnable parameters for sequence encoding.
We trained \abbrname for 10K steps by the Adam optimizer~\citep{kingma2014adam},
which took around 12/20 hours on 8 Nvidia V100 GPUs 
for code and AST, respectively.
For hyperparameter selections, 
we carefully aligned with our base models
as well as \citet{he2020momentum} regarding $\mathcal{L}_\text{DID}$ (\Eqref{eq:did}).
% After pre-training,
We evaluated \abbrname on four standard code understanding benchmarks
adopted by \citet{guo2022unixcoder}
including code-to-code search (abbreviated as code search)
on CodeNet,
clone detection on POJ-104~\citep{mou2016convolutional}, 
defect detection on Devign~\citep{zhou2019devign}
and text-to-code search (abbreviated as text search)
on CosQA~\citep{huang2021cosqa}.
We adopted mean average precision as the evaluation metric
for code search and clone detection,
accuracy for defect detection
and mean reciprocal rank for text search.
More details about our implementation 
and evaluation protocols can be found in Appendix~(\ref{ssec:appendix_details}). Code will be made publicly available.

% \subsection{Code Understandings and Generations}
\subsection{Effectiveness in Code Representation Learning}
\begin{table}[t]
\setlength{\tabcolsep}{0.01cm}
\centering
\caption{
Evaluations of code representations
% of the discrimination ability of code representations
on code search.
Results of our base models (CodeBERT and UniXcoder) 
are from~\citet{guo2022unixcoder}'s paper.
The column ``DYN'' indicates
whether a model was trained
using the test cases or not.
% of programs obtained with the help of fuzzing.
mAP scores (\%) are reported.
}
% Table generated by LaXcel from sheet 'laxcel'
\begin{tabular}{llclccclccclccclc}
\toprule
 &  &  &  & \multicolumn{3}{c}{Ruby} &  & \multicolumn{3}{c}{Python} &  & \multicolumn{3}{c}{Java} &  & \\ 
\cmidrule{5-7}\cmidrule{9-11}\cmidrule{13-15}
\multirow{-2}{*}{Model} &  & \multirow{-2}{*}{DYN} &  & Ruby & Python & Java &  & Ruby & Python & Java &  & Ruby & Python & Java &  & \multirow{-2}{*}{Overall}\\ 
\midrule
\textcolor[HTML]{AAAAAA}{CodeBERT} & \textcolor[HTML]{AAAAAA}{} & \textcolor[HTML]{AAAAAA}{\xmark} & \textcolor[HTML]{AAAAAA}{} & \textcolor[HTML]{AAAAAA}{13.55} & \textcolor[HTML]{AAAAAA}{3.18} & \textcolor[HTML]{AAAAAA}{0.71} & \textcolor[HTML]{AAAAAA}{} & \textcolor[HTML]{AAAAAA}{3.12} & \textcolor[HTML]{AAAAAA}{14.39} & \textcolor[HTML]{AAAAAA}{0.96} & \textcolor[HTML]{AAAAAA}{} & \textcolor[HTML]{AAAAAA}{0.55} & \textcolor[HTML]{AAAAAA}{0.42} & \textcolor[HTML]{AAAAAA}{7.62} & \textcolor[HTML]{AAAAAA}{} & \textcolor[HTML]{AAAAAA}{4.94}\\ 
CodeBERT-MLM &  & \xmark &  & 22.45 & 5.67 & 1.95 &  & 6.74 & 25.70 & 5.01 &  & 3.61 & 5.84 & 13.45 &  & 10.05\\ 
CodeBERT-MLM+RTD &  & \xmark &  & 13.22 & 1.00 & 0.10 &  & 1.24 & 14.35 & 1.20 &  & 0.20 & 0.18 & 6.34 &  & 4.20\\ 
\textbf{FuzzCodeBERT} & \textbf{} & \cmark & \textbf{} & \textbf{27.92} & \textbf{14.88} & \textbf{7.92} & \textbf{} & \textbf{15.39} & \textbf{30.47} & \textbf{10.26} & \textbf{} & \textbf{9.94} & \textbf{10.65} & \textbf{17.75} & \textbf{} & \textbf{16.13}\\ 
$\quad$- w/o DIM & \textbf{} & \cmark & \textbf{} & 24.05 & 14.08 & 6.96 & \textbf{} & 16.32 & 27.51 & 9.54 & \textbf{} & 8.66 & 9.76 & 13.49 & \textbf{} & 14.49\\ 
$\quad$- w/o DID & \textbf{} & \cmark & \textbf{} & 18.21 & 2.92 & 0.72 &  & 2.88 & 25.67 & 3.13 &  & 0.80 & 1.98 & 17.98 &  & 8.25\\ 
\midrule\midrule
\textcolor[HTML]{AAAAAA}{UniXcoder} & \textcolor[HTML]{AAAAAA}{} & \textcolor[HTML]{AAAAAA}{\xmark} & \textcolor[HTML]{AAAAAA}{} & \textcolor[HTML]{AAAAAA}{29.05} & \textcolor[HTML]{AAAAAA}{26.36} & \textcolor[HTML]{AAAAAA}{15.16} & \textcolor[HTML]{AAAAAA}{} & \textcolor[HTML]{AAAAAA}{23.96} & \textcolor[HTML]{AAAAAA}{30.15} & \textcolor[HTML]{AAAAAA}{15.07} & \textcolor[HTML]{AAAAAA}{} & \textcolor[HTML]{AAAAAA}{13.61} & \textcolor[HTML]{AAAAAA}{14.53} & \textcolor[HTML]{AAAAAA}{16.12} & \textcolor[HTML]{AAAAAA}{} & \textcolor[HTML]{AAAAAA}{20.45}\\ 
UniXcoder-MLM &  & \xmark &  & 20.49 & 13.54 & 3.25 &  & 10.40 & 19.49 & 3.69 &  & 4.13 & 5.14 & 12.29 &  & 10.27\\ 
UniXcoder-MLM+Contrast &  & \xmark &  & 30.83 & 25.73 & 16.46 &  & 25.44 & 30.50 & 16.80 &  & 16.01 & 17.26 & 18.86 &  & 21.99\\ 
\textbf{FuzzUniXcoder} & \textbf{} & \cmark & \textbf{} & \textbf{42.84} & \textbf{29.83} & \textbf{17.70} & \textbf{} & \textbf{33.73} & \textbf{47.77} & \textbf{21.94} & \textbf{} & \textbf{20.83} & \textbf{23.52} & \textbf{33.78} & \textbf{} & \textbf{30.22}\\ 
$\quad$- w/o DIM & \textbf{} & \cmark & \textbf{} & 22.50 & 13.52 & 6.66 & \textbf{} & 15.31 & 22.99 & 6.81 & \textbf{} & 7.54 & 6.84 & 12.94 & \textbf{} & 12.79\\ 
$\quad$- w/o DID & \textbf{} & \cmark & \textbf{} & 12.92 & 5.10 & 1.36 &  & 5.56 & 14.86 & 0.87 &  & 0.96 & 0.50 & 6.81 &  & 5.44\\ 
\bottomrule
\end{tabular}
\label{tab:code_search}
\end{table}

\paragraph{Learning with modality discrepancy.}
% As the dynamic program information (\ie, test cases)
% is often not available in downstream tasks in practice,
% the first question we studied here
% is whether the inconsistency of data modality
% between pre-training and deployment
% will refrain \abbrname from benefiting general code understanding.
As our \abbrname model aims to benefit
code representation learning with dynamic program information
which is often not available in downstream tasks in practice,
the first question to be studied here
is whether the inconsistency of data modality
between pre-training and deployment
will refrain \abbrname from benefiting general code understanding.
To that end,
we adopted the code search task
to identify equivalent functions
without fine-tuning the code representations
or learning additional classifiers.
Since our \abbrname models are trained on different data from the base models
(CodeBERT and UniXcoder),
we built several fairer baselines
to be trained under the exactly same settings as \abbrname
but learning from only
code or AST without test cases.
We presented CodeBERT/UniXcoder-MLM
to train by MLM solely as the baselines
following~\citet{liu2023contrabert},
% Moreover, 
% we constructed 
and CodeBERT-MLM+RTD/UniXcoder-MLM+Contrast to
adopt all the losses dedicated to code understanding in their papers
for comprehensive exploration on static information modeling.
% Note that UniXcoder didn't open-source their pre-training implementation,
% so we adapted from their predecessor~\citep{gao-etal-2021-simcse}
% to jointly learn by MLM and contrastive learning.
% To be concrete,
% we initalized the two static baselines 
% \textit{CodeBERT-MLM} and \textit{UniXcoder-MLM}
% by the weights of CodeBERT and UniXcoder in respective
% and train them by MLM for static information modeling (\Eqref{eq:sim}).
Correspondingly,
we denote the two variants of \abbrname
built upon CodeBERT and UniXcoder
as \textit{FuzzCodeBERT} and \textit{FuzzUniXcoder},
respectively.
% We also provided the results of the models
% trained without DIM or DID
% to investigate their necessity
% in dynamic information modeling.

As shown in Table~\ref{tab:code_search},
the superior performances attained by 
FuzzCodeBERT and FuzzUniXcoder
over their static baselines
demonstrate that 
\abbrname is able to yield discriminative code representations
that are beneficial to downstream tasks where
test cases
are not given.
% Besides,
Although Ruby is not one of the programming languages we used for pre-training,
\abbrname's improvements on such an unseen language
are still on par with that on Python and Java.
These results imply that
our models did learn to explore code semantics
rather than over-fitting to the training data.
We attribute the performance superiority obtained by \abbrname 
to the designs of not only modeling the dynamic information
jointly from code and test cases
but also distilling such knowledge
to be encoded into the feature representations of code.
This is evident by
the degradation of \abbrname
when training without either of the proposed components.
Such performance drops further verify
the effectiveness of our delicate designs
and demonstrate that
it is non-trivial to benefit code representation learning
by dynamic program information.
% the results of the \abbrname models 
% trained without DID %(FuzzCodeBERT/FuzzUniXcoder w/o DID),
% whose inferiority to the baselines
% indicate that the modality discrepancies
% between pre-training and inference
% can lead to dramatic model degradation
% if not properly handled.
% This further demonstrates 
% that it is non-trivial to benefit code representation learning
% by dynamic program information.
% regardless of its intuitiveness.

%
\begin{wrapfigure}{r}{0.5\textwidth}
\vspace{-0.4cm}
\centering
\includegraphics[width=1.\linewidth]{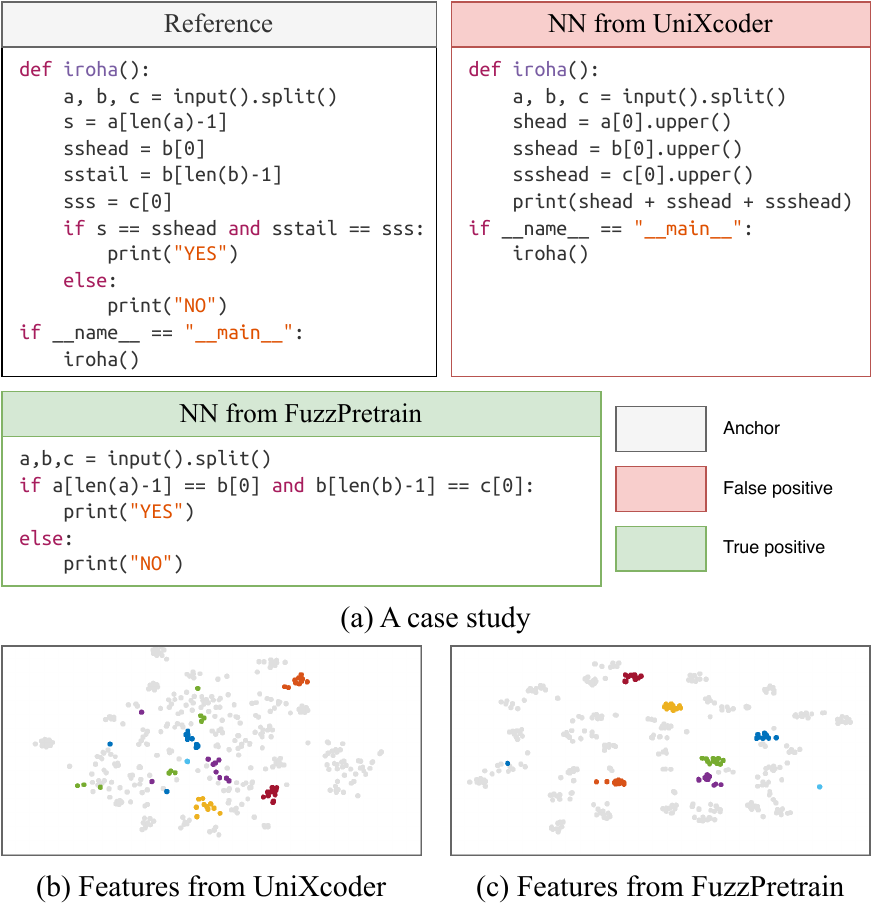} \vskip -0.1in
\caption{Qualitative studies for code search.
%on CodeNet~\citep{puri2021codenet}.
% Examples of code snippets on CodeNet~\citep{puri2021codenet}.
% The pairwise cosine similarity computed by different models are provided.
% The cosine similarity between code pairs computed by 
% either static-based model (\ie, UniXcoder~\citep{guo2022unixcoder})
% or our \abbrname is reported.
}
\label{fig:qualitative}
\vspace{-0.2cm}
\end{wrapfigure}
\paragraph{Qualitative studies.}
For more intuitive understanding of \abbrname's advantages
on code search,
we show an example in \Figref{fig:qualitative} (a)
to exhibit the nearest neighbors of a reference code snippet
decided by either UniXcoder or its \abbrname counterpart.
It is not surprising that
the code with similar structure (\eg variable or function names and the main entry point) 
can be easily confused with each other by the static-based method
even though the false positive example is with different purposes
from the reference.
On the other hand,
such functionality-wise relationships between code snippets 
are exposed by their execution behaviors,
hence,
are well captured by \abbrname.
% Besides,
To provide a global picture of the learned code features,
we adopted t-SNE~\citep{van2008visualizing} to visualize
the python code for 50 randomly selected problems (classes),
% from the data of code search,
which were encoded by the static-based model or our \abbrname
in \Figref{fig:qualitative} (b) and (c), respectively.
The functional equivalence of code 
are highlighted by the same colors,
% shared by different dots,
with only a few random problems are marked with bright colors to avoid chaos.
As depicted,
the features of functionally equivalent code snippets
yielded by UniXcoder
can sometimes spread over the feature space sparsely
due to their implementation variations.
However, 
our \abbrname forms more compact clusters
which are consistent with the underlying semantics of code.
These visualizations help explain
the potential benefits of jointly learning from
the static and dynamic information
to comprehensive code understanding.
% the effectiveness of jointly learning from
% the static and dynamic information
% for comprehensive code understanding.

\paragraph{Code understanding in novel domains.}
We empirically show on Table~\ref{tab:code_search} that 
addressing modality discrepancies between pre-training and inference
is critical to leveraging programs execution for learning code representation.
The next question we investigated is
whether our learned code features 
% encoding dynamic program information
are transferable and beneficial to downstream tasks in unseen data domains.
To that end,
we evaluated \abbrname on 
several unseen code understanding benchmark datasets~\citep{lu2021codexglue}.
% which involve different data from the CodeNet we adopted for pre-training.
Comparing FuzzCodeBERT/FuzzUniXcoder with the other methods in Table~\ref{tab:code_understanding_base},
we see non-negligible performance advantages 
% of \abbrname 
obtained by our \abbrname
% obtained by \abbrname in most cases
% on both POJ-104 and BigCloneBench
over these static-based methods 
% (CodeBERT-MLM/UniXcoder-MLM)
which learn from only the code structures.
% shown in Table~\ref{tab:code_understanding}
% of \abbrname.
The empirical advantage is particularly obvious for CodeBERT
which was initially less generalizable, 
since it learns from code by representing it as a sequence of tokens and
overlooks the structure information discovered 
by static analysis. %and available in syntactic representations.
% since it learns from ``plain'' code rather than its syntactic representations.
% and present its structure more ambiguously.
For UniXcoder, although introducing contrastive learning by
feeding the same code inputs to the encoder twice~\citep{gao-etal-2021-simcse} (\ie, ``Contrast'' in Table~\ref{tab:code_understanding_base})
is helpful to UniXcoder-MLM on defect detection, it leads to subtle performance degradation on the other two tasks. 
In fact, \abbrname can 
obtain a similar improvement (from 64.5\% to \textbf{65.6\%}) on the task of defect detection by
integrating such a code-to-code contrast (\ie, ``Contrast'') as well.
The effectiveness of combining ``Contrast'' with our \abbrname also implies the effectiveness of our dynamic information modeling on more advanced base models.
% into \abbrname
% However,
% % the data augmentation realized by dropout noise
% % is not guaranteed to be invariant to code semantics,
% % which leads to the uncertain effects of such 
% % a design to different downstream tasks.
% how to mutate code while maintaining its semantics
% to facilitate effective contrastive learning
% and benefit different downstream tasks universally
% remains an open question~\citep{liu2023contrabert}.
%
\begin{table}[t]
\parbox{.48\linewidth}{\footnotesize\setlength{\tabcolsep}{0.02cm}
\centering
\caption{
Comparisons with the static baselines %on code understanding downstream tasks
in novel data domains.
% The column ``Dynamic'' indicates whether test cases were used for training.
Results marked with $^*$ are reproduced using the checkpoints from authors
while the rest results are
directly collected from their papers or~\citet{guo2022unixcoder}'s paper.
}
% Table generated by LaXcel from sheet 'laxcel'
\begin{tabular}{llclclclc}
\toprule
Model &  & DYN &  & Clone & \multicolumn{1}{c}{} & Defect & \multicolumn{1}{c}{} & Text\\ 
\midrule
\textcolor[HTML]{AAAAAA}{CodeBERT} & \textcolor[HTML]{AAAAAA}{} & \textcolor[HTML]{AAAAAA}{\xmark} & \textcolor[HTML]{AAAAAA}{} & \textcolor[HTML]{AAAAAA}{82.7} & \textcolor[HTML]{AAAAAA}{} & \textcolor[HTML]{AAAAAA}{62.1} & \textcolor[HTML]{AAAAAA}{} & \textcolor[HTML]{AAAAAA}{65.7}\\ 
CodeBERT-MLM &  & \xmark &  & 88.7 &  & 63.5 &  & 67.4\\ 
CodeBERT-MLM+RTD &  & \xmark &  & 84.7 &  & 62.0 &  & 66.3\\ 
\textbf{FuzzCodeBERT} &  & \cmark &  & \textbf{93.0} &  & \textbf{64.1} &  & \textbf{69.1}\\ 
\midrule\midrule
\textcolor[HTML]{AAAAAA}{UniXcoder} & \textcolor[HTML]{AAAAAA}{} & \textcolor[HTML]{AAAAAA}{\xmark} & \textcolor[HTML]{AAAAAA}{} & \textcolor[HTML]{AAAAAA}{90.5} & \textcolor[HTML]{AAAAAA}{} & \textcolor[HTML]{AAAAAA}{64.5$^*$} & \textcolor[HTML]{AAAAAA}{} & \textcolor[HTML]{AAAAAA}{70.1}\\ 
UniXcoder-MLM &  & \xmark &  & 91.2 &  & 63.8 &  & 69.8\\ 
UniXcoder-MLM+Contrast &  & \xmark &  & 91.1 &  & \textbf{65.2} &  & 69.7\\ 
\textbf{FuzzUniXcoder} &  & \cmark &  & \textbf{92.2} &  & 64.5 &  & \textbf{70.7}\\ 
\bottomrule
\end{tabular}
\label{tab:code_understanding_base}
}\hfill
\parbox{.48\linewidth}{\footnotesize\setlength{\tabcolsep}{0.08cm}
\centering
\caption{
Comparisons with more state-of-the-arts that 
% were pre-trained on a similar amount of data.
are composed of similar number of learnable parameters.
% on code understanding.
Results marked with $^*$ are reproduced using the checkpoints from authors
while the rest results are directly collected from their papers or~\citet{guo2022unixcoder}'s paper.
}
% Table generated by LaXcel from sheet 'laxcel'
\begin{tabular}{llclclc}
\toprule
Model (Year) &  & Clone & \multicolumn{1}{c}{} & Defect & \multicolumn{1}{c}{} & Text\\ 
\midrule
RoBERTa (\citeyear{liu2019roberta}) &  & 76.7 &  & 61.0 &  & 60.3\\ 
GraphCodeBERT (\citeyear{guo2020graphcodebert}) &  & 85.2 &  & 62.9 &  & 68.4\\ 
DISCO (\citeyear{ding-etal-2022-towards}) &  & 82.8 &  & 63.8 &  & -\\ 
CodeRetriever (\citeyear{li2022coderetriever}) &  & 88.8 &  & - &  & 69.7\\ 
ContraBERT (\citeyear{liu2023contrabert}) &  & 90.5 &  & 64.2 &  & 66.7$^*$\\ 
CodeExecutor (\citeyear{liu2023code}) &  & 70.5$^*$ &  & 59.0$^*$ &  & 13.1$^*$\\ 
\textbf{FuzzCodeBERT} &  & \textbf{93.0} &  & 64.1 &  & 69.1\\ 
\textbf{FuzzUniXcoder} &  & 92.2 &  & \textbf{64.5} & \textbf{} & \textbf{70.7}\\ 
\bottomrule
\end{tabular}
\label{tab:code_understanding_sota}
} \vskip -0.1in
\end{table}

\paragraph{Comparisons with more state-of-the-arts.}
Although \abbrname adopted different pre-training data
from the popular bi-modal dataset~\citep{husain2019codesearchnet}
to enable compilation and fuzzing,
we compared it with the state-of-the-art models regardless to
demonstrate its effectiveness on code understanding.
% demonstrate the potential advantages of 
% learning with complements from
% program executions.
Specifically,
we compared \abbrname with three types of methods.
RoBERTa~\citep{liu2019roberta} learns at the natural language conventions.
DISCO~\citep{ding-etal-2022-towards},
CodeRetriever~\citep{li2022coderetriever},
and ContraBERT~\citep{liu2023contrabert}
benefit from contrastive learning as in our solution.
GraphCodeBERT~\citep{guo2020graphcodebert} and
CodeExecutor~\citep{liu2023code}
implicitly explore functionality
from DFG or execution traces, respectively.
Note that,
GraphCodeBERT extracts DFG from code on every downstream task
while CodeExecutor was evaluated without execution traces
to better align with practical application scenarios.
As shown in Table~\ref{tab:code_understanding_sota}, the performance advantages of \abbrname over GraphCodeBERT
implies that mining the functionality of programs
from the intricate dependencies among variables
is more challenging than 
modeling from the concrete input-output behavior
represented by test cases.
Besides,
CodeExecutor's inferiority compared to RoBERTa 
shows that it is difficult to adapt to downstream tasks when 
execution traces are not available.
% CodeExecutor was even inferior to RoBERTa,
% which
% % the inferiority of CodeExecutor
% % to RoBERTa
% reveals the obstacle to 
% benefit from it
% when execution traces are not available in practice.
Whilst the methods that are based on contrastive learning of 
code or its syntactic representations
yielded promising results,
\abbrname's superiority
demonstrate the effectiveness of learning with complements from 
dynamic information.
More importantly,
\abbrname can be integrated into those methods
to further benefit from more advanced modeling of static information.

\subsection{Components analysis and Ablation study}
\label{ssec:analysis}
\begin{figure}[t]
\centering
\begin{minipage}{0.32\textwidth}
\centering
\includegraphics[width=1.\textwidth]{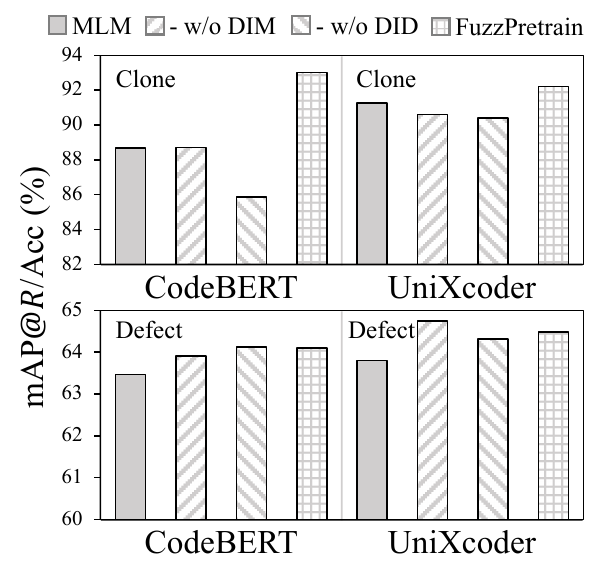} \vskip -0.2in
\caption{Effects of different components for dynamic information modeling. 
We constructed three variants of \abbrname
with either DIM or DID or both being removed 
to be compared.}
\label{fig:abl_task}
\end{minipage}\hfill
\begin{minipage}{0.32\textwidth}
\centering
\includegraphics[width=1.\textwidth]{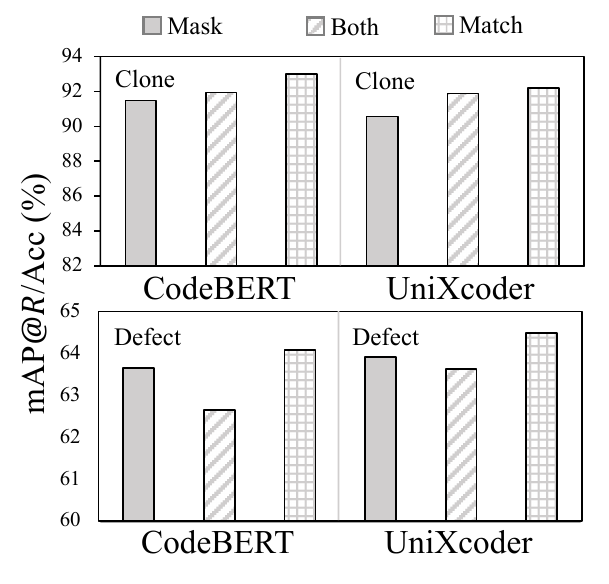} \vskip -0.2in
\caption{Dynamic information modeling by MLM.
% Justification of dynamic information matching (DIM).
The ``Mask'' variant replaces DIM by MLM for both code and test cases
while ``Match'' is the design we adopted
and ``Both'' is the combination of the two.
}
% We built an alternate of DIM (Match) 
% to model dynamic program information
% from test cases by masked tokens predictions (Mask).}
\label{fig:abl_dim}
\end{minipage}\hfill
\begin{minipage}{0.32\textwidth}
\centering
\includegraphics[width=1.\textwidth]{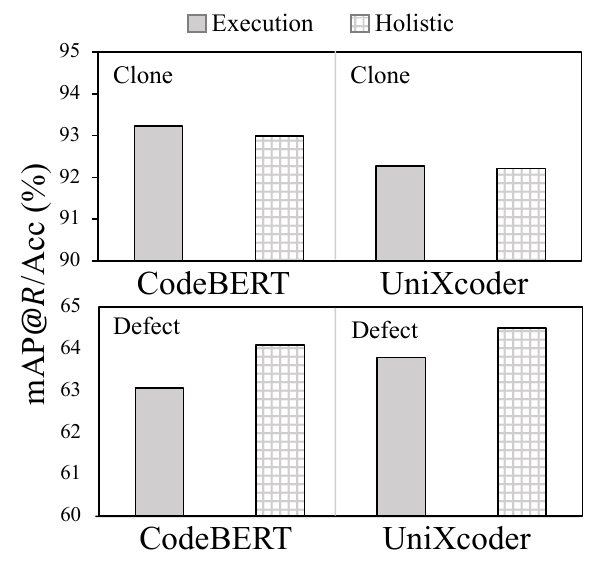} \vskip -0.2in
\caption{
Positive pairs in DID.
% Justification of dynamic information distillation (DID).
The ``Execution'' variant constructs the
positive pairs in DID using code $T^s$ and its test cases $T^d$,
and our ``Holistic'' design contrasts code to 
its concatenation with test cases $T^s\oplus T^d$.
% We conducted knowledge distillation from
% only the test cases (Execution)
% or the holistic representations of both the code and test cases.
}
\label{fig:abl_did}
\end{minipage}
\end{figure}

% We conducted comprehensive ablation studies
% on code understanding downstream tasks
% to deliver in-depth analyses and component justifications.
\paragraph{Effects of different components for dynamic information modeling.}
To effectively leverage the dynamic program information
encoded in test cases,
we proposed two components to 
derive such information and distill
it to enrich code representations by 
DIM (\Eqref{eq:dim}) and DID (\Eqref{eq:did}).
To study their independent contributions
to dynamic information modeling,
we constructed and compared three variants of \abbrname
by removing either or both of the two components.
% We studied their independent contributions
% to dynamic information modeling
% by constructing and comparing three variants of \abbrname
% by removing either or both of them.
As shown in \Figref{fig:abl_task},
the variant of \abbrname trained with only DID (w/o DIM)
often out-performed the baselines (MLM) trained with neither DIM nor DID.
% as well as the counterparts trained with only the DIM (w/o DID).
This indicates that the test cases concatenated 
after the source code or its syntactic representations
potentially play the roles of data augmentation
to perturb the distributions of code
by supplementing the dynamic information from test cases.
% and facilitate contrastive learning
% of discriminative representations,
% even though the dynamic information might not be fully explored
% in the absence of DIM.
Besides,
the static baselines trained without test cases
are always worse than
their counterparts trained with only DIM
on defect detection.
This supports our motivation that
the subtle changes or defects in code
can be revealed by their unexpected execution behaviors.
% the changes or defects in code
% can sometimes be unnoticeable
% in structure inspections
% but clearly revealed by their unexpected execution behaviors.
Moreover,
the generality of our \abbrname models on
both the retrieval (clone detection) and classification (defect detection)
tasks demonstrate the benefits of combining the two dynamic information modeling designs.
% On the other hand,
% the counterpart trained with only the DIM (w/o DID)
% is usually worse than the static baselines,
% implying that fitting the dynamic information about functionality
% can be distracting
% % to understanding the static information about structure
% when the source of information (test cases) is missing.
% By combining both the DIM and DID,
% our \abbrname model yielded the best results in most cases,
% which demonstrates the necessity of both the components. 
% to modeling dynamic program information. %from test cases.

\paragraph{Dynamic information modeling using MLM.}
Instead of modeling the dynamic program information %from test cases
by predicting the correspondence of code and test cases,
% we investigate the versatility of MLM for modeling both static and dynamic information
% by masked token predictions.
% To that end, we built two variants of \abbrname.
another intuitive idea is to
learn such knowledge in the same way as static information modeling,
\ie, by masked token predictions (MLM).
To investigate the versatility of MLM
for modeling both types of information,
we built two variants of \abbrname.
The ``Mask'' variant in \Figref{fig:abl_dim}
% denotes the variant which 
replaces DIM by MLM
for both code and test cases
while the ``Match'' is the DIM design we adopted
% (with MLM for code only)
and the ``Both'' combines the two. 
The consistent advantages of ``Match'' over ``Mask''/``Both''
indicates that applying MLM in test cases
is less effective in dynamic information modeling than DIM.
From our training logs, we also observe that
the encoder was able to accurately reconstruct the masked tokens in test cases (or code)
regardless of whether the code (or the test cases) is available in the model input.
This implies 
that syntactic representations and functional representations
may be well reconstructed independently, considering that their remaining tokens are still very informative, which makes it less straightforward to benefit one from the other using MLM.
On the contrary,
the labels for our DIM is defined only by the relationships between code and test cases, 
hence, it is infeasible to predict such labels 
without learning their correlations.
Besides,
considering the distribution discrepancies between
the code and test cases
(\eg test cases are likely to involve an exhaustive list of random numbers as inputs which are barely seen in code),
% syntactic and functional representations,
the code-agnostic knowledge about test cases
% learned by MLM
can even be distracting to code understanding
as demonstrated by the inferior performances yielded by the ``Both'' alternative.

\paragraph{Positive pairs in DID.}
We formulate DID to distill from the holistic representations
encoding both the static and dynamic information
as the latter is often not available in practice.
Given that code and its syntactic representations exist all the time,
it is natural to ask can we distill 
only the missing dynamic information
from test cases instead of the 
holistic information derived from also the code structure?
To address this concern,
we built a variant of \abbrname which 
formulates the $\mathcal{L}_\text{DIM}$
to identify test cases $T^d$
according to their corresponding code $T^s$ or AST
by constructing the positive pairs in \Eqref{eq:did}
to be $(T^s, T^d)$.
% constructs the positive pairs in contrastive learning
% by code/AST and their test cases ($T^s$, $T^d$),
% while that of our \abbrname is ($T^s$, $T=\{T^s\oplus T^d\}$).
We denote \abbrname as ``Holistic'' and its variant as ``Execution''
and compare them in \Figref{fig:abl_did}.
% and report their performance on downstream tasks in \Figref{fig:abl_did}.
Although the performances of the ``Execution'' variant
on clone detection
are on par with that of the ``Holistic'' counterpart,
its inferiority on defect detection
is non-negligible.
One potential solution to this problem
will be encoding
the two different modality of data
(\ie, code snippets and test cases)
by independent encoders
as commonly adopted in
cross-modal modeling methods
like vision-language models~\citep{radford2021clip,li2022blip}.
However,
to maintain the training efficiency and 
learning from the correlations between
the two modalities,
we stick to the single-encoder design in our \abbrname.

\section{Conclusion}
In this paper,
we make the first attempt
to derive the program functionality
from the dynamic information collected during executions
and leverage such knowledge
as the complements to the static structure information of code,
in order to facilitate comprehensive program profiling
and effective code representation learning.
% on leanring the functional semantics of programs
% as the complements to the structural semantics
% derived from their different static manifestations
% to facilitate
% comprehensive program understanding
% and effective code representation learning.
%
To that end,
we take advantage of fuzzing
% which is a primary dynamic code analysis technique,
to generate diverse and sufficient test cases 
for a large-scale code corpus efficiently
as explicit indications of programs' runtime behaviors.
To benefit from such a new modality of data
that is often not available on downstream tasks,
we proposed
\abbrname for joint static and dynamic information modeling.
Specifically,
\abbrname is trained
not only to accomplish
the conventional masked tokens predictions objective
but also
to learn the input-output relationships from test cases 
encoding the program-specific runtime behaviors,
as well as 
enforcing the model to infer such dynamic knowledge from code structures solely.
Extensive experiments
were conducted on various code understanding downstream tasks.
The notable performance advantages yielded by \abbrname
over the models learned from only the structure of code
demonstrate the potential benefits of the complements from program executions.
% given that the two pretext tasks designed for dynamic information modeling
% are both about the holistic comprehension of code in a global picture,
% we investigated but failed to observe the benefits of \abbrname
% to code generation downstream tasks usually conducted in tokens level.
% This inspires future studies on 
% benefiting token-wise code representation learning
% by dynamic program information.

\bibliography{conf_trans,egbib}
\bibliographystyle{iclr2024_conference}

\appendix
\section{Appendix}

\subsection{Implementation, Datasets and Evaluation protocols}
\label{ssec:appendix_details}
\paragraph{Datasets.}
Considering the non-negligible time consumption
of fuzzing,
we collected
1.2M code submissions
to $1000/1400/800/250$ problems
for C/C++/Python/Java
from CodeNet~\citep{puri2021codenet}
to be fuzzed as the training data of \abbrname.
We then evaluated \abbrname extensively
on four code understanding benchmark datasets:
(1) another subset of \textbf{CodeNet}~\citep{puri2021codenet}
collected by~\citet{guo2022unixcoder}
consisting of 50K functions
implemented in Python, Java, and Ruby
for solving one of $4,053$ online coding problems;
(2) \textbf{POJ-104}~\citep{mou2016convolutional} 
which contains $104$ C/C++ coding problems
with $500$ code submissions to each;
% (3) \textbf{BCB} is short for BigCloneBench~\citep{svajlenko2014towards}.
% We adopted its variant introduced by~\citet{lu2021codexglue},
% which includes
% 901K/416K/416K code snippets from ten different functionalities
% for training/validation/testing;
(3) \textbf{Devign}~\citep{zhou2019devign} which is composed of vulnerable functions
from four large and popular C-language open-source projects
with manual labels;
(4) \textbf{CosQA} which 
contains 20,604 pairs of code and 
real-world web queries~\citep{huang2021cosqa}
with annotations from human experts indicating
whether the questions raised by the queries can be properly addressed by the code.
All the data used for fine-tuning and testing
are carefully aligned with previous studies~\citep{feng-etal-2020-codebert,guo2022unixcoder}.
% and the training and evaluation subsets of CodeNet were carefully inspected
% to be disjoint from each other regarding the coding problems.

\paragraph{Evaluation protocols.}
We first investigated 
the discrimination ability of
the learned code representations
by code-to-code search 
(abbreviated as \textit{code search} in the paper)
on the subset of CodeNet collected by~\citet{guo2022unixcoder}.
In this task, submissions of the same coding problems
are assumed to share the same semantics
regardless of their implementations.
The feature distances between code pairs
were adopted to measure their semantic similarity
and the mean average precision (\textbf{mAP})
was reported
to quantify the quality of the retrieval results.
We then studied the effects of \abbrname
to several downstream tasks 
in unseen domains,
including clone detection, defect detection
and text-to-code search (abbreviated as \textit{text search}).
The objective of clone detection
is similar to that of code search, and we followed the same 
protocol of ~\citet{feng-etal-2020-codebert}'s work 
to test on POJ-104 and use 
% to identify
% semantically equivalent programs.
% Its difference to code search is that
% the pre-trained features are fine-tuned
% on POJ-104 following~\citet{feng-etal-2020-codebert}'s work.
\textbf{mAP}@$R$
to assess the results, with
% on POJ-104,
only the top-$R$ ($R=499$) most similar samples were considered in retrieval.
% We also explored
% the effects of solely improving code representations
% to text search
% by retrieving code snippets according to textual queries.
% in a similar way as on POJ-104.
In the task of text search, which requires retrieving code snippets according to textual queries, the mean reciprocal rank (\textbf{MRR})
is adopted as the metric following~\citet{guo2022unixcoder}'s work.
This evaluation was conducted on CosQA. % rather than
% CodeSearchNet~\citep{husain2019codesearchnet}
% because the latter had been used for pre-training our base models,
% and it is hard to tell whether the better performances
% were coming from more universal understanding of text-to-code alignments
% or more preservation of knowledge specific to the training domain.
Defect detection was carried out on Devign
and the accuracy (\textbf{Acc}) of binary classification is adopted
with a fixed threshold of $0.5$.

\paragraph{Implementation details.}
% We investigated the complementary effects of dynamic program information
% to either the source code or its syntactic representations (AST)
% by continue training CodeBERT~\citep{feng-etal-2020-codebert} and 
% UniXcoder~\citep{guo2020graphcodebert}.
Both our base models,
\ie CodeBERT~\citep{feng-etal-2020-codebert} and UniXcoder~\citep{guo2022unixcoder},
followed \citet{liu2019roberta} 
to take a 12-layers transformer
with 125M learnable parameters
% with 12 attention heads 
for sequence encoding.
We followed their designs to 
set the batch size to $2048$ and $1024$
while the maximum sequence length
to $512$ and $1024$
% set both the batch size and the maximum sequence length to $512$ and $1024$ 
for CodeBERT and UniXcoder, respectively.
In inputs, $400$ and $800$ tokens are reserved for code and AST, respectively, 
and the rest are for test cases.
% We set the maximal length of input sequences greater when using AST
% because it generally contains 70\% more tokens than code~\citep{guo2022unixcoder}.
The test cases of each program
were concatenated with the code or the AST by the separation token
until reaching the length limits, while the rest was dropped.
% \todo{statistic of test cases, number of cases per program, average length of test cases}.
The \abbrname model was updated by the Adam optimizer~\citep{kingma2014adam} during training
with a learning rate of $2e-5$ for $10K$ steps.
For dynamic information distillation $\mathcal{L}_\text{DID}$ (\Eqref{eq:did}),
we followed~\citet{he2020momentum}
to set the momentum coefficient $m=0.999$,
the temperature $t=0.07$,
and the number of random samples $l^n=2^{16}$.
% We also followed~\citep{lample2019xlm,guo2020graphcodebert,guo2022unixcoder}
% to alternate among different pretext tasks
% at each training steps.
% We also followed~\citep{guo2020graphcodebert,guo2022unixcoder}
% to refrain the models from biasing to frequent programming languages on the training data
% by constructing each batch
% from a random language selected from a multinomial distribution
% $q_i={p_i^\alpha}/{\sum_{j=1}^n p_j^\alpha}$
% where $p_i$ is the frequency of the $i$-th language and $\alpha=0.7$
% is a smoothen factor.
% Specially,
% we observed severe forgetting problems in continual learning
% so combined the original data used by baselines (CodeSearchNet)
% with the new fuzzed code corpus (CodeNet)
% then subsampled it by $10\times$
% to alleviate forgetting and accelerate training.
The overall pre-training process took round 12/20 hours
on 8 Nvidia V100 GPUs
for training with code and AST, respectively.

% \subsection{Further explorations of baselines}

% In this appendix,
% we provide additional experiments
% to conduct more comprehensive analyses of \abbrname
% and also discuss its limitations
% that are potentially inspiring to future works.
\subsection{Limitations}
\paragraph{Fuzzing code corpus.}
Our current pre-training data is restricted to 
OJ-like code corpus (\ie, CodeNet)~\citep{puri2021codenet},
which refrains us from ablating affecting factors in the data distribution in making fair comparison to existing methods.
To be more specific, most commonly adopted code corpus~\citep{husain2019codesearchnet}
are composed of standalone functions spread over
various software projects (\eg, CodeSearchNet),
whose test cases cannot be easily obtained. 
% which are not directly compilable and cannot be fuzzed.
% In this case,
% we carried out pre-training on CodeNet~\citep{puri2021codenet}
% that comes with oracle for compilation.
Whilst 
OJ data is showing some unique characteristics
to benefit our \abbrname model 
on understanding similar code snippets
as indicated by our remarkable performance advantages on POJ-104~\citep{mou2016convolutional}
(Table~\ref{tab:code_understanding_sota}),
this also limits our model's generalization ability to 
other type of code corpus,
% Therefore,
% we occasionally observed performance drops of \abbrname 
% in comparisons to the base models,
\eg the F1-score of clone detection on BigCloneBench~\citep{svajlenko2014towards} 
yielded by our FuzzUniXcoder
was $1\%$ lower than that by UniXcoder pre-trained on CodeSearchNet. Yet, when both pre-trained on the same selected subset of CodeNet, our \abbrname leads to +0.9\% F1 gain in comparison to existing pre-training strategies using, for example, the MLM loss on CodeBERT. 
% Moreover,
Exploring fuzzing on more diverse code corpus help address this limitation.

\paragraph{Text-code tasks.}
Following the discussion about fuzzing code corpus in the previous paragraph, we would like to mention that, since CodeNet does not contain text description of each code, pre-training on it may not fully unleash the power of pre-training on text-code downstream tasks. That is to say, although we have shown the effectiveness of our \abbrname on the text code search task in Tables~\ref{tab:code_understanding_base} and~\ref{tab:code_understanding_sota}, even better results can be obtained if we can pair the CodeNet data with text descriptions or if we can pre-train on a dataset with not only texts and code but also test cases.
This also withholds FuzzCodeBERT and FuzzUniXcoder from
surpassing every state-of-the-art methods on
text-code tasks.
% Nonetheless,
In addition to exploiting datasets, extensive experiments presented in this paper
also verifies complementary effects of dynamic program modeling to these methods, which implies that combining more advanced methods~\citep{wang2023codet5,li2022soft} with our \abbrname also leads to superior performance than that of FuzzCodeBERT and FuzzUniXcoder.
% We believe those stronger static-based state-of-the-arts
% can also benefit from our \abbrname
% upon fixing the issue on fuzzing arbitrary code snippets.
% evidence the effectiveness of learning from dynamic program information
% as the complements to two representative static-based methods
% learned from either source code or its AST.
% By fixing the issue on fuzzing standalone functions,
% we believe
% the methods with more advanced designs on static information modeling
% can also be benefited by the dynamic information obtained during executions.

\paragraph{Code generation.}
Our designs for dynamic information modeling
are all about the holistic comprehensions of code in a global picture,
while how to benefit token-wise code understanding by using it 
is not straightforward.
Indeed,
we did not observe 
the improvements brought by \abbrname
on code generation tasks
which are usually conducted at token-level,
leaving an interesting problem to be studied in the future. 

\end{document}